# Making Tennis Fairer: The Grand Tiebreaker

Steven J. Brams

Department of Politics

New York University

New York, NY  10012

USA

steven.brams@nyu.edu

Mehmet S. Ismail

Department of Political Economy

King's College London

London WC2R 2LS

UK

mehmet.ismail@kcl.ac.uk

D. Marc Kilgour

Department of Mathematics

Wilfrid Laurier University

Waterloo, Ontario N2L 3C5

CANADA

mkilgour@wlu.ca

**Abstract**

Tennis, like other games and sports, is governed by rules, including the rules that determine the winner of points, games, sets, and matches. If the two players are equally skilled—each has the same probability of winning a point when serving or when receiving—we show that each has an equal chance of winning games, sets, and matches, whether or not sets go to a tiebreak. However, in a women's match that is decided by 2 out of 3 sets, and a men's match that is decided by 3 out of 5 sets, it is possible that the player who wins the most games may not be the player who wins the match. We calculate the probability that this happens and show that it has actually occurred—most notably, in the 2019 men's Wimbledon final between Novak Djokovic and Roger Federer, which took almost five hours to complete and is considered one of the greatest tennis matches ever (Djokovic won).

We argue that the discrepancy between the game winner and the match winner, when it occurs, should be resolved by a Grand Tiebreak (GT)—played according to the rules of tiebreaks in sets—because each player has a valid claim to being called the rightful winner. A GT would have the salutary effect of giving each player an incentive to strive hard to win every game—even every point—lest he/she win in sets but lose more games. This would make competition keener throughout a match and probably decrease the need for a GT, because the game and set winner would more likely coincide when the players fight hard for every point.

# 1. Introduction

In Kemeny and Snell (1960), the authors use Markov chains to calculate the probability that players with different probabilities of winning points go on to win games, sets, and matches. Subsequently, there have been numerous studies to expand on this work, which now include tiebreaks in sets (Pollard, 1983; Haigh, 1996; MacPhee et al., 2004; Brams and Ismail, 2018; Brams et al., 2018). The rules for tiebreaks, which did not exist in 1960, have varied but have now been standardized in the four so-called Grand Slam Tournaments, which we assume here.

There is a consensus that serving in a game provides the server with an advantage, which is mitigated by the rule that servers alternate over the games of a set. Still, if P serves first in a set and Q second, with alternation thereafter, P has a better chance of winning if the set ends at 6-1 or 6-3, because P serves in one more game, whereas at the other set scores without a tiebreaks (6-0, 6-2, 6-4, and 7-5), the even number of games ensures that P and Q serve first in the same number of games, giving them equal chances of winning the set if they are equally skilled.

But Haigh (1996) argued that serving first in a set does not advantage the first server, and MacPhee et al. (2004) generalized this argument to a broader class of contests. Our calculations support this conclusion. Haigh's key observation was that extending a set a certain number of games beyond the point when a winning score is reached does not affect the result, because the winning player will remain ahead by the required winning margin. For example, if the set ends with a score of 6-3, if Q is given another game to serve, P will still win the set by the required margin of two games even if Q wins the extra game.

In this paper, we focus on the discrepancy between the match winner, who wins more sets, and the winner of the most games. Although both the theoretical probability of this discrepancy's happening (about 5 percent) and the empirical probability that it actually has occurred in the four Grand Slam Tournaments through 2024 are low (less than 2 percent). Nonetheless, the discrepancy between a game winner and a set winner recently occurred in one Grand Slam men's final (Wimbledon, 2019), demonstrating how the choice of a champion in the most storied tennis tournament in the world would have changed if the criterion for winning the tournament had been the game winner rather than the set winner.

To settle who is the rightful winner when there is such a discrepancy—the game or set winner—we recommend that there be a Grand Tiebreak (GT), played according to the rules of tiebreaks in sets. A GT would have the salutary effect of giving each player an incentive to strive hard to win every game—even every point—lest he/she lose in sets after winning more games. This would make competition keener throughout a match and probably decrease the need for a GT, because the two winners would more likely coincide when they are fighting for every point.

## 2. Markov Chains and Games

A tennis match comprises several distinct contests—games, sets, and the match itself—that may have different winners. A set consists of a sequence of games, with possibly a tiebreak, and a match consists of a sequence of sets. In men's Grand Slam matches, the winner is the first player to win 3 out of 5 sets; in women's, the winner is the first player to win 2 out of 3 sets.

A coin toss determines which player starts the match by serving in the first game. Let $p \in (0,1)$ be $A$'s scoring (i.e., point-winning) probability when $A$ serves, and $q \in (0,1)$ be $B$'s scoring probability when $B$ serves. Only one player serves in each game, but who does so strictly alternates between games. A player who first wins at least 6 games by a margin of at least 2 games wins the set. Thus, if $A$ wins, the score in games is $6 - 0, 6 - 1, 6 - 2, 6 - 3, 6 -$ 4, or 7 – 5; if *B* wins, these scores are reversed. If neither player meets these conditions, the set must be tied at 6 games each. Then it goes to a tiebreak, which is won by the first player to win two more points than his/her opponent after one player wins at least 6 points (more on the tiebreak later).

If *A* serves in a set, we assume that *A* wins each point with probability *p* and loses it with probability 1 – *p*. Throughout the paper, we assume that serves are independent events. As Figure 1 illustrates, we can model a tennis game as a Markov chain, with states distinguished according to score.

Starting at state 0 – 0 at the bottom of the chain, each serve moves the game upward on the chain until it reaches one of the five states on the top row of Figure 1, where the states are labelled by *A*'s *relative score* (*A*'s total points won minus *B*'s total points won). We call state (*A*, 2) a *direct win* by A ("Game *A*"), and state (A, –2) a direct win by *B* ("Game *B*"). (*Indirect wins* occur when there is a tiebreak before one player wins, which we will come to later.)

By direct calculation, the probabilities of reaching state $S = (A, r)$ are:

$$P(A, 2) = p^4(5 - 4p), P(A, 1) = 4p^3(1 - p)^2, P(A, 0) = 6p^2(1 - p)^2,$$

$$P(A, -1) = 4p^2(1 - p)^3, \text{ and } P(A, -2) = (1 - p)^4(1 + 4p)$$

If the game is tied 40 – 40 (called "deuce"), *A*'s relative score is 0. After *A* serves, *A*'s relative score increases by 1 with probability *p* and decreases by 1 with probability 1 – *p*. If *A*'s relative score reaches 2 (after his/her tennis score is greater than *B*'s, before or after deuce, by at least 2 points), *A* wins a game. If *A*'s relative score reaches –2, *B* wins. Thus, the states with relative scores 2 and –2 are absorbing states. The three non-absorbing states correspond to *A*'s relative scores of +1, 0, and –1.

In the tiebreak, if $S$ is a non-absorbing (non-terminal) state, let $P(A, S)$ represent the probability that $A$ wins eventually, given that his/her current position is $S$. Using standard methods, we can calculate $P(A, r)$, the probability that $A$ wins eventually given that A is about to serve in a (game) tiebreak in which $A$'s relative score is $r$:

$$P(A, 1) = \frac{p - p^2 + p^3}{1 - 2p + 2p^2}, P(A, 0) = \frac{p^2}{1 - 2p + 2p^2}, P(A, -1) = \frac{p^3}{1 - 2p + 2p^2}$$

Note that the probability that $B$ wins from state $S$ is always $1 - P(S)$. For instance, the probability that $B$ wins the game starting with relative score $r = 0$, with $A$ to serve, of course, is $1 - P(A, 0) = \frac{1 - 2p + p^2}{1 - 2p + 2p^2} = \frac{(1-p)^2}{1 - 2p + 2p^2}$.

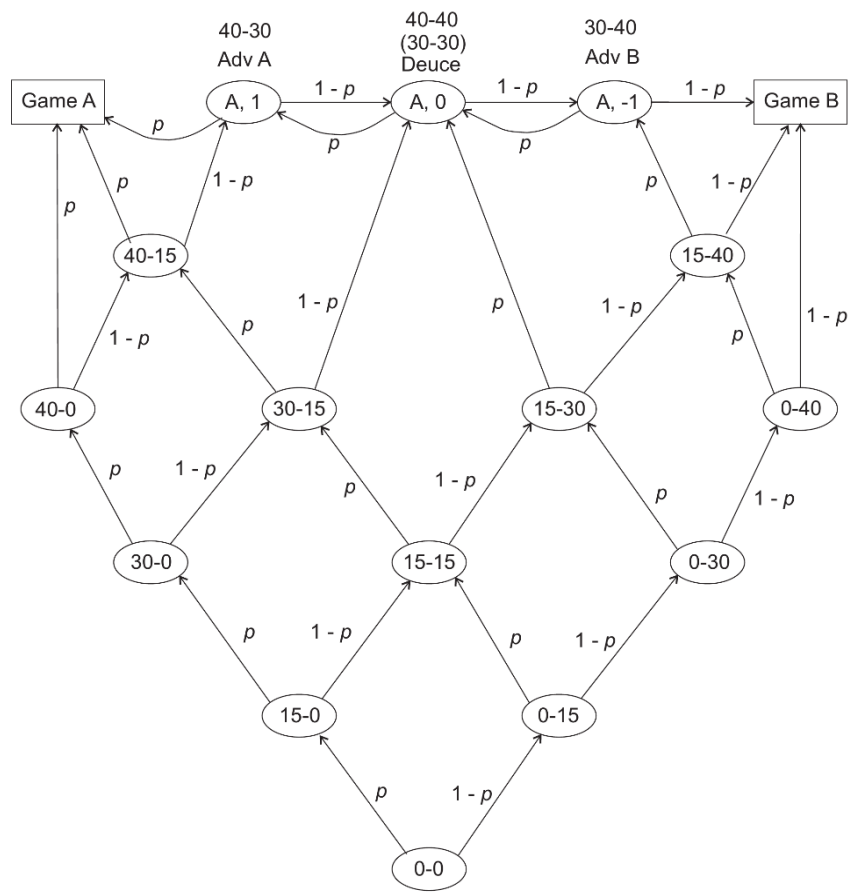

**Figure 1. Markov chain when A serves first**

Combining this information, we find that $A$'s probability of winning a game, starting at state $0 - 0$ when $A$ is serving, is

$$\frac{p^4(15 - 34p + 28p^2 - 8p^3)}{1 - 2p + 2p^2} \tag{1}$$

The probability that $B$ wins the game is the complementary probability, which can be obtained by replacing $p$ by $q$.

### 3. Tiebreak of a Set

As noted earlier, each set in a tennis match is a sequence of games with alternating servers. But the rules change if the set is tied in games at 6-6. A tiebreak of a set involves serves by both players in a specific order. After $A$'s first serve, $B$ serves twice in a row, then $A$ serves twice in a row, and so on. This is commonly known as the ABBA rule, because the service order can be represented by the following sequence: A/BB/AA/BB/AA….

A regular tiebreak of a set, called a 7-point tiebreak, requires that the winner obtain at least 7 points and win by a margin of at least 2 points. In particular, if there is a 6-6 tie, the tiebreak goes to "sudden death," whereby the first player to go ahead by 2 points wins.[1] If $A$ wins before sudden death, the *tiebreak score* is (7, $x$), where $x = 0, 1, 2, 3, 4,$ or 5. Conversely, if $B$ wins before sudden death, the tiebreak score is ($x$, 7), where $x = 0, 1, 2, 3, 4,$ or 5.

---

[1] Since 2022, if the last set in a Grand Slam Tournament is the 5th for men or the 3rd for women, and a tiebreak is required at 6-6 in that set, a 10-point rather than a 7-point tiebreak is played to determine the winner. This change has little effect on our subsequent results, so we do not include it in our calculations but instead assume a 7-point tiebreak throughout the match.

**Table 1. Probability of tiebreak scores (7, *x*) or (*x*, 7), where *x* = 0, 1, 2, 3, 4, or 5.**

$$P(7,0) = p^3(1-q)^4$$

$$P(0,7) = (1-p)^3 q^4$$

$$P(7,1) = \sum_{y=3}^{4} \binom{3}{y-1}\binom{4}{7-y} p^y(1-p)^{4-y} q^{y-3}(1-q)^{7-y}$$

$$P(1,7) = \sum_{y=3}^{4} \binom{4}{y}\binom{3}{6-y} p^{y-3}(1-p)^{7-y} q^y(1-q)^{4-y}$$

$$P(7,2) = \sum_{y=3}^{5} \binom{4}{y-1}\binom{4}{7-y} p^y(1-p)^{5-y} q^{y-3}(1-q)^{7-y}$$

$$P(2,7) = \sum_{y=2}^{4} \binom{4}{y}\binom{4}{6-y} p^{y-2}(1-p)^{7-y} q^y(1-q)^{4-y}$$

$$P(7,3) = \sum_{y=2}^{5} \binom{5}{y}\binom{4}{6-y} p^y(1-p)^{5-y} q^{y-2}(1-q)^{7-y}$$

$$P(3,7) = \sum_{y=2}^{5} \binom{4}{y-1}\binom{5}{7-y} p^{y-2}(1-p)^{7-y} q^y(1-q)^{5-y}$$

$$P(7,4) = \sum_{y=1}^{5} \binom{5}{y}\binom{5}{6-y} p^y(1-p)^{5-y} q^{y-1}(1-q)^{7-y}$$

$$P(4,7) = \sum_{y=2}^{6} \binom{5}{y-1}\binom{5}{7-y} p^{y-2}(1-p)^{7-y} q^y(1-q)^{6-y}$$

$$P(7,5) = \sum_{y=1}^{6} \binom{5}{y-1}\binom{6}{7-y} p^y(1-p)^{6-y} q^{y-1}(1-q)^{7-y}$$

$$P(5,7) = \sum_{y=1}^{6} \binom{6}{y}\binom{5}{6-y} p^{y-1}(1-p)^{7-y} q^y(1-q)^{6-y}$$

Let $P(7, x)$ and $P(x, 7)$, where $x = 0, 1, 2, 3, 4,$ or $5$, denote the probabilities that the tiebreak score is $(7, x)$ and $(x, 7)$, respectively. Table 1 presents the respective formulas for these probabilities. In each formula of Table 1, the quantity $y$ represents the number of points won by the winner on the winner's own serves. For $P(7, 0)$, $y = 3$; for $P(0, 7)$, $y = 4$; for all other probabilities in the table, there are at least two possible values of $y$, so it can be used as the index of summation.

The probability that a tiebreak continues—so that the winner in fact scores more than 7 points—which we call the "tiebreak of the tiebreak," is

$$P(6, 6) = \sum_{y=0}^{6} \binom{6}{y} \binom{6}{6-y} p^{6-y}(1-p)^y q^{6-y}(1-q)^y$$

Finally, the unconditional probabilities that $A$ and $B$ win in a tiebreak are given by the following formulas:

$$P^A(\text{tiebreak}) = P(6, 6) \left( \frac{p - pq}{p + q - 2pq} \right)$$

$$P^B(\text{tiebreak}) = P(6, 6) \left( 1 - \frac{p - pq}{p + q - 2pq} \right) = P(6, 6) \left( \frac{q - pq}{p + q - 2pq} \right)$$

We next explain where the term $\frac{p-pq}{p+q-2pq}$ comes from.

**Tiebreak of a Tiebreak**

Consider a 7-point tiebreak that reaches a score of $(6, 6)$. This tiebreak is resolved by an additional tiebreak modeled by the Markov subchain shown in Figure 2.

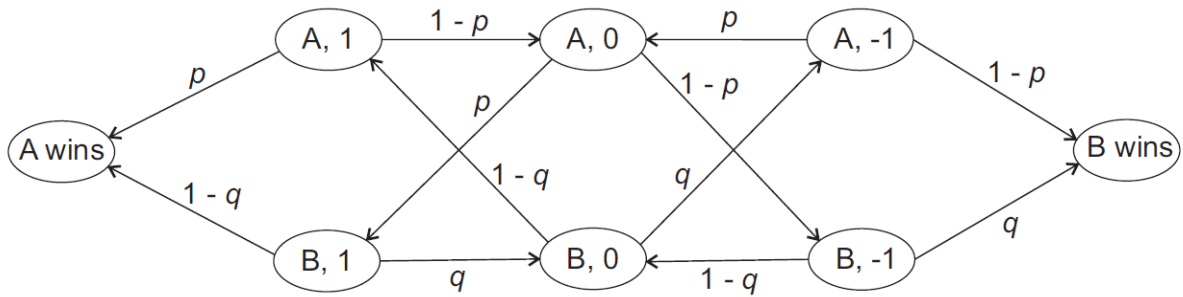

**Figure 2. Markov subchain for tiebreak of a tiebreak**

This subchain consists of eight states, two of which are absorbing, one with a win for *A* and the other with a win for *B*. The non-absorbing states are labelled according to which player has the next serve, and the value of *r*, the relative advantage of player *A*.

If the tiebreak reaches the score 6-6, *A*'s relative advantage is 0, and the tiebreak of the tiebreak begins in state (*A*, 0), with *A* serving. The outcome of *A*'s serve is either a win for *A* (with probability *p*) or a loss (with probability $1 - p$). At the next state, $r = 1$ if *A* won, or $r = -1$ if *A* lost. In either case, the next player to serve is player *B*. Therefore, the next state will be either (*B*, 1) or (*B*, –1). Note that *B*'s serve results in a win for *B* (which is a loss for *A*) with probability *q*, or a win for *A* (which is a loss for *B*) with probability $1 - q$. It can be checked that the order of serving is either AB or ABBA or ABBAAB or ABBAABBA …, continuing until one player has two wins more than the other—that is, until *A*'s relative advantage is either $r = 2$ or $r = -2$.

If *S* is a non-absorbing state, let $P_S$ represent the conditional probability that *A* wins eventually, given that his/her current position is *S*. Of course, if the current position is *S*, the probability that *B* wins eventually is $P_S^B = 1 - P_S$.

We calculate these probabilities, beginning with the states with $r = -1$. Clearly,

$P_{A,-1} = p\,P_{A,0}$ and $P_{B,-1} = (1-q)P_{B,0}$

Next,

$$P_{A,0} = p\,P_{B,1} + (1-p)P_{B,-1} = p\,P_{B,1} + (1-p)(1-q)P_{B,0}$$

$$P_{B,0} = (1-q)\,P_{A,1} + q\,P_{A,-1} = (1-q)\,P_{A,1} + q\,p\,P_{A,0}$$

But also

$$P_{A,1} = p + (1-p)P_{A,0} \text{ and } P_{B,1} = (1-q) + q\,P_{B,0}$$

Substitution now produces

$$P_{A,0} = p\,(1-q) + p\,q\,P_{B,0} + (1-p)(1-q)P_{B,0} = p\,(1-q) + [p\,q +$$
$$(1-p)(1-q)]P_{B,0}$$

which we write as

$$P_{A,0} = p\,(1-q) + \Delta\,P_{B,0}$$

where $\Delta = p\,q + (1-p)\,(1-q)$. By a parallel process, we find that

$$P_{B,0} = p\,(1-q) + \Delta P_{A,0}$$

We can solve for $P(A, 0)$ by substituting to obtain

$$P_{A,0} = p\,(1-q) + \Delta\left[p\,(1-q) + \Delta\,P_{A,0}\right] = p\,(1-q)(1+\Delta) + \Delta^2 P_{A,0}.$$

It follows that

$$P_{A,0} = \frac{p(1-q)(1+\Delta)}{1-\Delta^2} = \frac{p(1-q)}{1-\Delta} = \frac{p - pq}{p + q - 2pq}$$

Of course, the probability that $B$ wins eventually, given the process starts at state $S = (A, 0)$, is

$$P_{A,1}^{B} = 1 - P_{A,0} = 1 - \frac{p - pq}{p + q - 2pq} = \frac{q - pq}{p + q - 2pq}$$

It can be checked that $P_{B,0} = P_{A,0}$. In other words, the probabilities are unchanged if the initial server is $B$ as long as the service pattern $(1, 2, 2, …)$ is maintained.

## 4. Probability of Winning a Set

Let $\bar{p} \in (0,1)$ denote the probability that $A$ wins a game in which $A$ serves, and let $\bar{q} \in (0,1)$ denote the probability that $B$ wins a game in which $B$ serves. Note that $p$ and $q$ refer to probabilities of winning a point, whereas $\bar{p}$ and $\bar{q}$ refer to probabilities of winning a game, as shown in Equation (1) in Section 2. If $A$ wins directly (i.e., without a tiebreak), the *set score* is either $(7, 5)$ or $(6, x)$, where $x = 0, 1, 2, 3,$ or 4. If $B$ wins directly, the set score is either $(5, 7)$ or $(x, 6)$, where $x = 0, 1, 2, 3,$ or 4. Figure 3 illustrates a tennis set as a Markov chain.

The probability that the set score is $(6, x)$, denoted $\bar{P}(6, x)$, where $x = 0, 1, 2, 3,$ or 4, is given by

$$\bar{P}(6,x) = \begin{cases} \displaystyle\sum_{y=3-m}^{3+m} \binom{2+m}{y-1}\binom{3+m}{6-y} \bar{p}^{6-y}(1-\bar{p})^{y-3+m} \bar{q}^{3+m-y}(1-\bar{q})^y & \text{if } x = 2m \\ \displaystyle\sum_{y=2-m}^{3+m} \binom{3+m}{y}\binom{3+m}{5-y} \bar{p}^{6-y}(1-\bar{p})^{y-2+m} \bar{q}^{3+m-y}(1-\bar{q})^y & \text{if } x = 2m+1. \end{cases}$$

Note that the index of summation, $y$, equals the number of times that the winner "breaks serve," i.e., wins a game in which the opponent serves. Similarly, the probability that the set score is $(x, 6)$, denoted $\bar{P}(x, 6)$, where $x = 0, 1, 2, 3,$ or 4, is given by

$$\bar{P}(x,6) = \begin{cases} \displaystyle\sum_{y=3-m}^{3+m} \binom{3+m}{y}\binom{2+m}{5-y} \bar{p}^{3+m-y}(1-\bar{p})^y \bar{q}^{6-y}(1-\bar{q})^{y-3+m} & \text{if } x = 2m \\ \displaystyle\sum_{y=3-m}^{4+m} \binom{3+m}{y-1}\binom{3+m}{6-y} \bar{p}^{4+m-y}(1-\bar{p})^y \bar{q}^{6-y}(1-\bar{q})^{y-3+m} & \text{if } x = 2m+1. \end{cases}$$

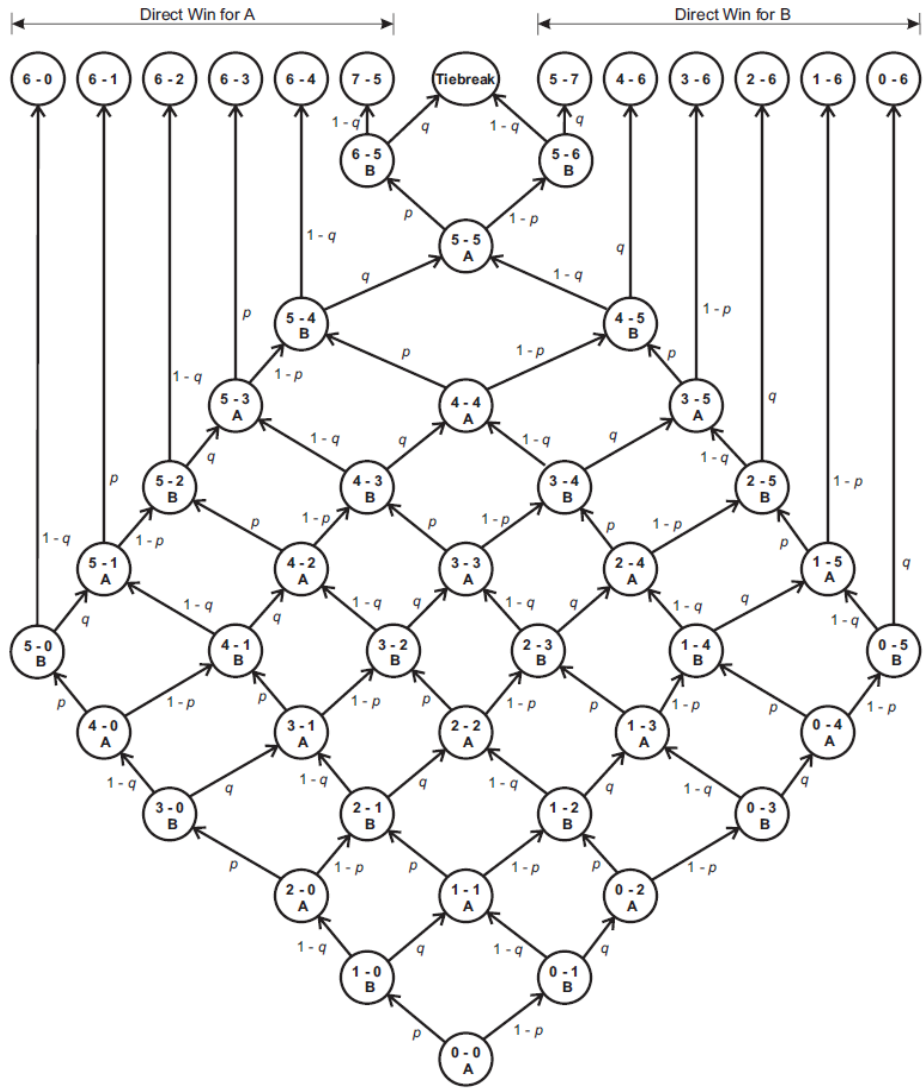

**Figure 3 Markov chain of a tennis set**

The probabilities of set scores $(7, 5)$ and $(5, 7)$ are

$$\bar{P}(7,5) = \sum_{y=0}^{5} \binom{5}{y}\binom{5}{5-y} \bar{p}^{\,6-y}(1-\bar{p})^{y}\bar{q}^{\,5-y}(1-\bar{q})^{y+1},$$

$$\bar{P}(5,7) = \sum_{y=0}^{5} \binom{5}{y}\binom{5}{5-y} \bar{p}^{\,5-y}(1-\bar{p})^{y+1}\bar{q}^{\,6-y}(1-\bar{q})^{y}$$

The probability that the set goes to a tiebreak is

$$\bar{P}(6,6) = 1 - \sum_{x=0}^{4} \left( \bar{P}(6,x) + \bar{P}(x,6) \right) - \bar{P}(7,5) - \bar{P}(5,7)$$

If a set goes to tiebreak and $A$ wins, its score is recorded as (7, 6); if it goes to tiebreak and $B$ wins, the recorded score is (6, 7). Thus, the probabilities of set scores (7, 6) and (6, 7) are

$$\bar{P}(7,6) = \bar{P}(6,6)P^A(\text{tiebreak})$$

### 4. The Grand Tiebreak and the Data

As noted earlier, the winner of a match (i.e., by sets) may win strictly fewer games than his/her opponent. We proposed that a Grand Tiebreak (GT) be used in such a situation to determine the player who more deserves to win the match. In this section, we assess the frequency of this discrepancy's occurring both in theory and in a dataset of Grand Slam Tournaments.

**The Grand Tiebreak**

Men's and women's Grand Slam Tournaments are played as best-of-5 match and best-of-3 matches, respectively, so we focus on these two formats. In a best-of-$(2k+1)$ match, a *set sequence* is defined as a sequence of set scores in which one player wins $k+1$ sets out of (at most) $2k+1$ of the sets played in the match. For example, in a best-of-5 match, $A$ can win a match in three sets with a set sequence of [(6, 4), (6, 0), (7, 5)], or in four sets with a set sequence of [(6, 1), (2, 6), (7, 6), (7, 5)]. In a best-of-$(2k+1)$ match, where $k = 1$ or 2, if $A$ wins, the *match score* is $(k + 1, x)$, and if $B$ wins, the match score is $(x, k + 1)$, where $x = 0$, 1, …, $k$.

We begin by calculating the total number of possible match scores and then determine how many of these lead to a GT. We earlier showed that there are 7 different winning set scores for a player: (6, 0), (6, 1), (6, 2), (6, 3), (6, 4), (7, 5), and (7, 6).

First consider a best-of-3 match in women's Grand Slam Tournaments. We start by calculating the number of different set scores when the match concludes in 2 sets. Because there are 7 possible winning scores for each set, the number of different scores for a match that ends in two sets is $2 \times 7^2 = 98$. In this case, it is impossible for the winner to win fewer total games than the loser, because the winner must win more games in both sets.

There are two distinct cases in which a player can win with a (2, 1) or (1, 2) match score, because the winner can lose either the first or the second set (but not the third set). For each of these cases, a (2, 1) or (1, 2) match score can occur in $7^3 = 343$ ways. Because either player can win, it follows that there are $2 \times 2 \times 343 = 1372$ set scores in a match that lasts three sets. Analogous calculations for best-of-5 matches in men's Grand Slam Tournaments give a total of 686, 14,406, and 201,684 distinct match scores in matches that end in 3, 4, or 5 sets, respectively.

To determine the number of ways a GT can occur, we count the match scores in which the winner wins strictly fewer games, in total, than the loser. First, consider a best-of-3 match. For a GT to occur, the final score must be either (2, 1) or (1, 2). Using a computer search, we found that there are 136 distinct match scores that lead to a GT for matches that end in three sets. Because there are in total $98 + 1372 = 1470$ distinct match scores, the percentage of match scores with a GT is $\frac{136}{1470} \approx 9.25\%$.

Now consider a best-of-5 match. Similar calculations to those above show that there are 180 distinct match scores that lead to a GT for matches that end in 4 sets, and 32,124 distinct match scores that end in 5 sets. Overall, the percentage of match scores requiring a GT is

$$\frac{180 + 32{,}124}{686 + 14{,}406 + 201{,}684} = \frac{32{,}304}{216{,}776} \approx 14.90\%.$$

Thus, in almost 15 percent of all match scores, the winner wins fewer games, in total, than the loser.

The above calculations show that a non-negligible proportion of match scores leads to a GT if each set score is equally likely. But even casual observation suggests that a (7, 5) set score is much more probable than a (6, 0) sweep. This observation motivates the definition of a GT probability, which is the weighted average of match scores requiring a GT, with the weights based on the probability of each specific set score.

**The GT Probability**

Recall that $p$ is $A$'s probability of winning a point when he/she serves, and $q$ is $B$'s probability when he/she serves. For different values of $p = q$, we calculate, using a computer program, the *GT probability*—the probability that the match winner wins fewer total games than the loser.

**Table 2. GT probabilities for different equal values of *p* and *q***

|            | (1/2, 1/2) | (3/5, 3/5) | (3/4, 3/4) | (5/6, 5/6) |
|------------|------------|------------|------------|------------|
| Best-of-5  | 4.77%      | 5.12%      | 5.15%      | 2.37%      |
| Best-of-3  | 3.20%      | 3.31%      | 3.17%      | 1.62%      |

Notice that the best-of-5 GT probabilities are uniformly greater than the best-of-3 probabilities, indicating that longer matches raise the probability of a discrepancy between the game and the set winners when the players are equally skilled at serving. The maximum probabilities in each case are intermediate probabilities (3/4 for best-of-5, 3/5 for best-of-3), which are the two probabilities that best reflect the actual advantage of serving in Grand Slam Tournaments.

To illustrate the calculation of the GT probabilities in Table 2, consider the set sequence [(4, 6), (6, 0), (6, 0), (4, 6), (4, 6)], which leads to a GT. $B$ wins this match with 3 sets to *A*'s 2 sets, but *A* wins 24 games compared to *B*'s 18. According to the formulas in section 3, for $p = ½$, the probability of a (4, 6) set score is 0.1230, and the probability of a (6, 0) set score is 0.0156. Thus, because we treat each set score as independent, the probability of the set sequence [(4, 6), (6, 0), (6, 0), (4, 6), (4, 6)] is calculated as follows:

$$0.1230^3 \times 0.0156^2 = 4.529 \times 10^{-7}$$

with 3 identical "competitive" sets and 2 that are decidedly not. For best-of-5 when $(p, q)$ = (1/2, 1/2) in Table 2, the value 4.77% is obtained by summing the probabilities of all set sequences that require a GT in a best-of-5 match. Similarly, in a best-of-3 match, the GT probability is 3.20% for $(p, q)$ = (1/2, 1/2).

The GT probability varies depending on the value of $p \in (0,1)$. When $p$ tends toward 0 or 1, the GT probability approaches 0. The GT probability is maximized around $p \approx 0.7$, where it reaches a peak value of 5.7% in a best-of-5 match. The decrease in GT probability for values of $p$ exceeding 0.7 is clear. For example, when $p = \frac{5}{6}$, the GT probability decreases to 2.37%; when $p = 0.95$, it falls below 1%.

**Grand Slam Dataset**

We utilize a dataset compiled by Jeff Sackmann / TennisAbstract.com (available at [www.github.com/JeffSackmann](www.github.com/JeffSackmann)), which includes matches from 1968, the beginning of the Open Era in tennis, up to and including 25 August 2024, prior to the 2024 US Open Grand Slam Tournament. In total, the dataset comprises 340,156 professional singles matches (187,914 men's and 152,242 women's). Our primary focus is on the 49,901 completed matches (25,399 men's and 24,502 women's) in Grand Slam Tournaments (i.e., excluding matches which ended because one player retired).

Table 3 provides summary statistics for the Grand Slam Tournament dataset, including the total number of matches, sets, and games played. It also shows the number of matches where the winner won fewer games in total than the loser (referred to as *Empirical GTs*). The table compares these statistics across men's and women's Grand Slam matches, as well as those from the most recent three years (2022, 2023, and 2024, up to and including Wimbledon 2024).[2]

**Table 3. Grand Slam summary satistics and empirical GT probability percentages**

| | Matches | Empirical GTs | Empirical GT % | Best-of-5 GT Probability (%) $p = q = 0.64$ | Best-of-3 GT Probability (%) $p = q = 0.58$ |
|---|---|---|---|---|---|
| Men's | 25,399 | 484 | 1.91 | 5.41 | |
| Women's | 24,502 | 277 | 1.13 | | 3.28 |
| Men's (last 3 years) | 1,143 | 33 | 2.89 | 5.41 | |
| Women's (last 3 years) | 1,137 | 12 | 1.06 | | 3.28 |

The Empirical GT Percentage column in the table presents the actual percentage of GTs for each category. The empirical GT probability is less than the theoretical GT probability (for men, about 2 versus 5 percent). But we note that about 14 matches per Grand Slam tournament would require a GT, a number we consider significant.

The GT Probability columns show the theoretical GT probabilities for men's and women's Grand Slam matches. These probabilities are calculated using empirical server point-winning percentages: $p = q = 0.64$ for men and $p = q = 0.58$ for women, which reflects the stronger and more difficult-to-return serves of men. The data is based on matches from Wimbledon and the US Open in 2022, 2023, and 2024. The dataset includes 145,205 points from men's matches and 91,685 points from women's matches.

---

[2] Since 2022, all Grand Slam Tournaments have followed the tiebreak rules we describe in the text.

We think the gap between theoretical and empirical GT percentages is mainly because, early in a tournament, there are many matches between contestants who are not equally skilled. Higher-ranked players typically win most games and most sets, making a GT very unlikely. Straight-set wins are indeed about three percentage points higher in the first three rounds of Grand Slams than in the last three rounds: 49.4 percent vs. 46.7 percent for men, and 70.8 percent vs. 67.6 percent for women.

## 5. Conclusions

We analyzed men's and women's Grand Slam tennis matches under the present rules of play, assuming that the players were equally skilled at winning points. We calculated the probability that the set winner does not win the most games in both men's and women's matches (about 5 percent for men and 3 percent for women).

We compared these figures with the actual number of matches in which this difference arose in actual play and found the theoretical percentage to be much higher than the actual percentage (5 vs. 1.9 percent). We think this is mostly explained by the fact that, especially early in a tournament, the contestants are not equally skilled. In fact, the more highly rated players tend to win by more decisive scores in early matches, often in straight sets, in which a discrepancy between the set and the game winner cannot arise, making a GT moot.

But once the 128 contestants are whittled down to two in the first seven rounds of play, the two finalists are usually more or less equally skilled. The men's final match is very rarely won in only 3 sets. If, in addition, the game winner is different from the set winner, we think it only fair that this discrepancy be resolved by a GT.

Startlingly, this discrepancy occurred in the 2019 Wimbledon final between Novak Djokovic and Roger Federer, which Djokovic won in five sets that took almost five hours to complete. (This made it the longest Wimbledon final ever; after reaching 12 games all in the 5th set, Djokovic won the set tiebreak 7-3.) But Federer beat Djokovic in almost every statistic used to measure tennis performance, including winning more games (36 to 32).[3] A GT would have given Federer the opportunity to test Djokovic's mettle, which seems only fair to decide such a nail-biting contest.

A saving grace of GT is that it will almost surely diminish the likelihood that it happens, because players will have an incentive not to coast in a set that they are likely to lose. Competing for every game, and even for every point, in order to win as many games as possible, will make sets more engaging contests throughout a match.

The only downside to a GT, we think, is that it will be more exhausting for players, both physically and mentally. But shouldn't physical stamina and mental resilience be considered the hallmarks of skill in sports?

---

[3] https://en.wikipedia.org/wiki/2019_Wimbledon_Championships_%E2%80%93_Men%27s_singles

# References

Brams, Steven J, and Mehmet S Ismail. 2018. "Making the Rules of Sports Fairer." *SIAM Review* 60 (1): 181–202.

Brams, Steven J, Mehmet S Ismail, D Marc Kilgour, and Walter Stromquist. 2018. "Catch-Up: A Rule That Makes Service Sports More Competitive." *American Mathematical Monthly* 125 (9): 771–796.

Haigh, J. (1996). More on n-point, win-by-k games. *Journal of Applied Probability*, 33(2), 382-387.

Kemeny, John G., and J. Laurie Snell (1960; reprinted 1976). *Finite Markov Chains.* New York: Springer.

MacPhee, I. M., Rougier, J., and Pollard, G. H. (2004). Server advantage in tennis matches. *Journal of Applied Probability*, 41(4), 1182-1186.

Pollard, G. H. (1983). An analysis of classical and tie-breaker tennis. *Australian Journal of Statistics,* 25(3), 496-505.